\documentclass[12pt]{article}

\usepackage{amsmath}
\usepackage{amsfonts}
\usepackage{amssymb}

\begin{document}

\title{THE STANDARD MODEL AND THE GENERALIZED COVARIANT DERIVATIVE}

\author{M. Chaves and H. Morales\\
\it Escuela de F\'{\i}sica, Universidad de Costa Rica\\
\it San Jos\'e, Costa Rica\\
\it E-mails: mchaves@cariari.ucr.ac.cr, hmorales@ariel.efis.ucr.ac.cr
}

\date{June 25, 1999}

\maketitle
\begin{abstract}
The generalized covariant derivative, 
that uses both scalar and vector bosons, is defined. 
It is shown how a grand unified theory of the Standard
Model can be constructed using a generalized Yang-Mills 
theory.
\end{abstract}

\section{The generalized covariant derivative}
\noindent
Last year the authors developed a generalized 
Yang-Mills theory (GYMT) that used a covariant 
derivative that included not only vector bosons 
but scalar fields as well.~\cite{11}, \cite{13}
The motivation at the time was to simplify the 
writing of the multiple terms of the 
Glashow-Weinberg-Salam (GWS) model using $U(3)$. 
Our inspiration was an old idea by Fairlie and 
Ne'eman.~\cite{15}
The idea is that the Higgs 
bosons fit neatly in the adjoint of $SU(2/1)$, 
along with the gauge vector bosons, and that 
the hypercharges of the leptons are given 
correctly by one of the diagonal generators 
of that graded group.
This model has two main problems, reviewed 
by us.~\cite{11}
We then proposed
that the problems of the old model could be resolved 
if one switched to the $U(3)$ Lie group, since it 
is possible to obtain the correct quantum numbers for all 
the particles of the GWS {\it if instead of the usual 
Gell-Mann representation a 
different one is used}.
This new representation 
is a linear combination of generators of the 
usual one. In this model an extra scalar boson makes its 
appearance, but it decouples from all the other 
particles.

Here we study the application 
of a GYMT to the 
building of a grand unified theory (GUT) 
of the Standard Model 
at the rank 5 level. It turns out that, at this rank, 
there is only one possible GUT, and 
it is based on the group $SU(6)$.
The grand unification 
group has to contain two $SU(3)$'s, one to represent 
flavordynamics and another to represent chromodynamics. 
The algebra of $SU(6)$ has $SU(3)\otimes SU(3)\otimes U(1)$ 
as the group associated with one of its maximal 
subalgebras.
It turns out that this group gives, using 
a generalized covariant derivative, all the correct quantum 
numbers of all the fermions, vector bosons, 
GWS Higgs and GUT Higgs.
{\it Again, as in the GWS case, 
this does not happen in the usual representation, but 
in a different one that 
is a linear combination of the generators of other.} 
It is in this new representation that 
the particles of the Standard Model appear with 
their correct quantum numbers.

In what follows we will give a description of how the GUT is constructed 
and its overall structure. Certain
dynamical details are, for the moment, left out, since at this time
we are not finished with our calculations.

\section{Quick review of the generalized covariant derivative}
\noindent
We define the generalized covariant $D$ to transform as 
a four-vector contracted with Dirac matrices.
Assume we have an expression that is invariant under a Lorentz
transformation and contains a contracted 4-vector $\hbox{$\not \! \! A$}$.
If now we were to substitute this contracted vector by $\gamma^5\varphi$,
where $\varphi$ is a scalar field, then, from the properties of the 
Lorentz spinorial representation the expression would still be
Lorentz-invariant.

Consider a Lie group with $N$ generators.
Associate $N_V$ generators with an equal number of vector gauge 
fields $A_\mu^a$, and $N_S$ 
generators with an equal number of scalar gauge fields $\varphi^b$, 
with $N=N_V+N_S$.
Then the covariant derivative $D$ is defined by
\begin{equation}
D \equiv \hbox{$\not \! \partial$} +\hbox{$\not \! \! A$} + \Phi
\label{general}
\, , \end{equation}
with
\begin{eqnarray}
\hbox{$\not \! \! A$} & \equiv & \gamma^{\mu} A_\mu \equiv ig \gamma^{\mu} A_\mu^{\ a} T^a \, , \qquad 
\qquad a = 1, \ldots, N_V \, , \nonumber \\
\Phi    & \equiv & \gamma^5 \varphi \equiv -g \gamma^5 \varphi^b T^b \, , \qquad 
\qquad b = N_V+1, \ldots, N  \nonumber
\, . \end{eqnarray}
We take the gauge transformation for these fields to be
\begin{equation}
\hbox{$\not \! \! A$} + \Phi \to U (\hbox{$\not \! \! A$} + \Phi) U^{-1} - (\hbox{$\not \! \partial$} U) U^{-1}
\, , \end{equation}
therefore, we can have
\begin{equation}
D \to U D U^{-1}
\label{transform}
\, . \end{equation}
If the theory is going to contain fermions they must be placed in an 
irrep that is either the fundamental or at least can be constructed 
from products involving the fundamental or its conjugate, to assure 
gauge invariance.
This point will be illustrated later.
The non-abelian lagrangian is constructed based 
on the requirements that it should contain only fermion 
fields and covariant derivatives, and possess
both Lorentz and gauge invariance:
\begin{equation}
{\cal L}_{NA} = \overline{\psi} i D \psi + \frac{1}{2g^2} \widetilde{{\rm Tr}\,} 
\left( \frac{1}{8} {\rm Tr}\,^2 D^2 - \frac{1}{2} {\rm Tr}\, D^4 \right)
\label{lagNA}
\, , \end{equation}
where the trace with the tilde is over the Lie group matrices and the one
without it is over matrices of the spinorial representation of the Lorentz
group.
The additional factor of $1/2$ that the traces of (\ref{lagNA}) have 
comes from the usual normalization in the non-abelian case
\begin{equation}
\widetilde{{\rm Tr}\,}\, T^a T^b = \frac{1}{2} \delta_{ab}
\label{norma}
\, . \end{equation}

If we expand the covariant derivative into its component
fields,~\cite{11} the lagrangian (\ref{lagNA}) shows to be made of terms that are
traditional in Yang-Mills theories:
\begin{eqnarray}
{\cal L}_{NA} &=& \overline{\psi} i(\hbox{$\not \! \partial$} + \hbox{$\not \! \! A$}) \psi - g \overline{\psi} i 
\gamma^5 \varphi^b T^b \psi + \frac{1}{2 g^2} \widetilde{{\rm Tr}\,} \left( \partial_{\mu} A_
\nu - \partial_{\nu} A_\mu + [A_\mu, A_\nu] \right)^2 \nonumber \\
   & & + \frac{1}{g^2} \widetilde{{\rm Tr}\,} \left( \partial_{\mu} \varphi 
+ [A_{\mu}, \varphi] \right)^2
\label{lagexpandido}
\, , \end{eqnarray}
where the first term on the right looks like the usual matter term of a gauge theory,
the second like a Yukawa term,
the third like the kinetic energy of vector bosons in a Yang-Mills theory and
the fourth like the gauge-invariant kinetic energy of scalar 
bosons in the non-abelian adjoint representation. 

We call the differential operators in equations (\ref{transform})
and (\ref{lagNA}) {\it unrestrained},~\cite{11} because they 
keep acting indefinitely to the right.
However, in the expanded form of (\ref{lagexpandido}), 
after having done all the algebra, the differential operators there 
are {\it restrained}, that is, the partial derivatives acts only 
on the immediately succeeding functions to the right.

Notice that in GYMTs the gauge invariance is given by the full Lie group.
That is, the lagrangian is completely invariant under transformation 
(\ref{transform}).
This does not mean that the form of the transformed 
covariant derivative has to remain exactly the same.
Similarly, there are
transformations in the group that will mix chiralities. 
But after doing all the transformations the final result is invariant.
Interestingly enough, in both the case of the GWS model of Ref.~1 and 
the GUT theory studied here, the maximal subgroup maintains the chiralities
of the sectors of fermions, and the reason is very clear: it is a 
Yang-Mills theory. 
In other words, the GYMTs we have studied 
contain a typical Yang-Mills theory that uses as Lie group the 
maximal subgroup of the GYMT.

\section{The group generators}
\noindent
The choice of our unifying group is to be guided by the requirement that
it should contain $SU(3)_C \otimes U(3)$, where the ``C" stands for 
{\em color}. 
The $U(3)$ is necessary because it contains the GWS
model using GYMTs. 
Therefore the candidate group should be at least rank 5, 
(= rank 2 due to $SU(3)_C$ plus rank 3 due to $U(3)$.)
The smallest such group is $SU(6)$, which contains $SU(3) \otimes 
SU(3) \otimes U(1)$ as a maximal subgroup,~\cite{22}
so that the first $SU(3)$ is a color subgroup and 
$SU(3) \otimes U(1)$ can be identified with $U(3)$, because their
Lie algebras are the same and the connection between triality of color and
electric charge can be established through the $SU(3) \otimes U(1)$'s 
embedding in $SU(6)$ that unifies the coupling constants.

The $SU(6)$ diagonal generators in the
$SU(3)_C \otimes SU(3) \otimes U(1)$ decomposition (normalized by 
(\ref{norma}) are:
\begin{eqnarray}
T_3^C &=& \frac{1}{2} {\rm diag} (1,-1,0,0,0,0) \, . \nonumber \\
T_8^C &=& \frac{1}{2 \sqrt{3}} {\rm diag} (1,1,-2,0,0,0) \, . \nonumber \\
T_3 &=& \frac{1}{2} {\rm diag} (0,0,0,1,-1,0) \, . \nonumber \\
T_8 &=& \frac{1}{2 \sqrt{3}} {\rm diag} (0,0,0,1,1,-2) \, . \nonumber \\
T_{35} &=& \frac{1}{2 \sqrt{3}} {\rm diag} (1,1,1,-1,-1,-1) \, .
\label{generadores}
\end{eqnarray}
The two first generators are the diagonal ones of QCD, the third represents
the electrically neutral component of the $SU(2)$ that is included in the 
subgroup $SU(3)$
and the last two are related to the hypercharge as we shall soon see.

Similarly to what happens in our model for GWS, the two diagonal generators 
would seem to be the assignments of the isospin $T_3$ and the 
hypercharge $Y$, but this would give the wrong
value for the hypercharge of the Higgs boson.
To correct this problem we changed the group and used $U(3)$ instead.
With the help of the extra generator we could obtain correctly all the
quantum numbers of the GSW model through a linear combination of it
and the original hypercharge generator.
Consequently the two last generators in (\ref{generadores}) must be related
to the hypercharge and to the new scalar boson.

We shall rename the fourth $T_8$ in (\ref{generadores}) $T_{Y'}$, 
because it stands for the original
hypercharge in our $SU(3)$ electroweak model, and
the fifth $T_{35}$ as the generator $T_{Z'}$, because
it represents the extra one in our $U(3)$ model that permitted us 
to avoid the wrong hypercharge assignment of the Higgs boson. 
So we discover here the origin of the trick we had to do in our
first paper.

\section{The generators and the fermions quantum numbers}
\noindent
Take $SU(6)$ in the $SU(3)_C \otimes SU(2) \otimes U(1)$ 
decomposition, that will give us the quantum numbers of 
the particles in the large irreps.
The expected results are the 
quantum numbers of the fermions of the Standard Model
\begin{equation}
\begin{array}{rclcrcl}
L_e = (\nu\ e)_L^T &:& ({\bf 1},{\bf 2})_{1} & ,\ {\rm and}\  & L_u 
= (u\ d)_L^T &:& ({\bf 3},{\bf 2})_{-1/3} \\
e_R &:& ({\bf 1},{\bf 1})_{2} & & u_R &:& ({\bf 3},{\bf 1})_{-4/3} \\
\nu_R &:& ({\bf 1},{\bf 1})_{0} & & d_R &:& ({\bf 3},{\bf 1})_{2/3} \, , 
\end{array}
\label{fermions}
\end{equation}
where the ``T" stands for ``transposed". We are using the 
Gell-Mann-Nishijima relation in the form $Q=T_3 - \frac{1}{2} Y$.
We are looking for a $\bf 15$-dimensional representation to accommodate 
the fermions.

The branching rule~\cite{22} for this representation into the fundamental
ones of $SU(3)_C \otimes SU(3) \otimes U(1)_{Z'}$ is
\begin{equation}
{\bf 15} = ({\bf \bar{3}},{\bf 1})_{2} + ({\bf 1},{\bf \bar{3}})_{-2} + ({\bf 3},{\bf 3})_{0}
\, . \end{equation}
In these entries for the {\bf 15}, call them $({\bf x},{\bf y})_{Z'}$, 
generically, the $\bf x$ and $\bf y$ irreps belong to $SU(3)_C$ and 
$SU(3)$, respectively.
The subindex $Z'$ is the value of 
the $U(1)_{Z'}$ generator when acting on the states given by the irreps.
Therefore in the second term of this equation we expect to accommodate
the antileptons; thus we shall work with the 
$\bf \overline{15}$-dimensional representation.
We need to decompose the $SU(3)$ in terms of its maximal subgroup
$SU(2) \otimes U(1)_{Y'}$, in order to recognize the fermion fields in the
$SU(3)_C \otimes SU(3) \otimes U(1)_{Z'}$ branching rule.
Employing ${\bf 3} = ({\bf 2})_1 + ({\bf 1})_{-2}$ for $SU(3)$,
the $\bf \overline{15}$-dimensional representation of $SU(6)$ can be broken up
into irreps of $SU(3)_c \otimes SU(2) \otimes U(1)_{Y'} \otimes U(1)_{Z'}$:
\begin{equation}
{\bf \overline{15}} = ({\bf 3},{\bf 1})_{0,-2} + ({\bf 1},{\bf 2})_{1,2} + 
({\bf 1},{\bf 1})_{-2,2} + ({\bf \bar{3}},{\bf \bar{2}})_{-1,0} + ({\bf \bar{3}},{\bf 1})_{2,0}
\, . \end{equation}
Notice that neither $Y'$ nor $Z'$ represent the correct hypercharge for
$L_e$ and $e_R$ as given in (\ref{fermions}), but we use them to find the
correct linear combination that gives the hypercharge.

Take 
\begin{equation}
T_Y = \alpha T_{Y'} + \beta T_{Z'}
\, . \end{equation}
From the normalization condition we get the condition for the 
coefficients $\alpha$ and
$\beta$ in order that $T_Y$ is normalized directly by the above 
equation, i.e.
\begin{equation}
1 = \alpha^2 +\beta^2
\, . \end{equation}

From inspection, the correct combination to obtain the quantum 
numbers of the leptons is
\begin{equation}
T_{Y} = - \frac{1}{\sqrt{5}} T_{Y'} + \frac{2}{\sqrt{5}} T_{Z'}
\label{TY}
\, . \end{equation}
The accompanying generator $T_Z=\gamma T_{Y'} + \delta T_{Z'}$ 
is obtained from the orthogonality of the generators, with the 
result 
\begin{equation}
T_{Z} = \frac{2}{\sqrt{5}} T_{Y'} + \frac{1}{\sqrt{5}} T_{Z'}
\label{TZ}
\, . \end{equation}
Explicitly, these new generators become
\begin{eqnarray}
T_{Y} &=& \frac{1}{2} \sqrt{\frac{3}{5}} {\rm diag} \left(\frac{2}{3}, 
\frac{2}{3}, \frac{2}{3}, -1, -1, 0\right) \, . \nonumber \\
T_{Z} &=& \frac{1}{2} \sqrt{\frac{1}{15}} {\rm diag} \left( 1, 1, 1, 1, 1, -5\right)
\, . \end{eqnarray}
We shall denote them as the {\it hypercharge} and the {\it ultracharge}
generator, respectively.
All the other generators of $SU(6)$ are left unchanged.
From now on we use this new representation in all calculations.

Let us write the branching rule for the $\bf \overline{15}$-dimensional irrep
into irreps of $SU(3)_C \otimes SU(2) \otimes U(1)_Y \otimes U(1)_Z$, using
the values of $Y$ and $Z$ as subindices, in that order:
\begin{equation}
{\bf \overline{15}} = ({\bf 3},{\bf 1})_{-4/3,-1} + ({\bf 1},{\bf 2})_{1,2} 
+ ({\bf 1},{\bf 1})_{2,-1} + ({\bf \bar{3}},{\bf \bar{2}})_{1/3,-1} + ({\bf \bar{3}},{\bf 1})_{-2/3,2}
\, . \end{equation}
The last two terms are the antiquarks.
Explicitly, the ${\bf \overline{15}}$, which is the antisymmetric tensor
product of two fundamentals, can be written as 
\begin{equation}
\psi = \frac{1}{\sqrt{2}} \left( \begin{array}{rrr|rr|r} 
0         & u_R^3     & -u_R^2    & -d_R^{1c} & u_R^{1c} & d_L^{1c} \\
-u_R^3    & 0         & u_R^1     & -d_R^{2c} & u_R^{1c} & d_L^{2c} \\
u_R^2     & -u_R^1    & 0         & -d_R^{3c} & u_R^{1c} & d_L^{3c} \\ \hline
d_R^{1c}  & d_R^{2c}  & d_R^{3c}  & 0         & -e_R     & \nu_L    \\
-u_R^{1c} & -u_R^{2c} & -u_R^{3c} & e_R       & 0        & e_L      \\ \hline
-d_L^{1c} & -d_L^{2c} & -d_L^{3c} & -\nu_L    & -e_L     & 0        \\
\end{array} \right)
\, . \end{equation}
The quark colors have been denoted 1, 2 and 3.
The identification of \mbox{$(d$ $-u)^T$} as a $\bf \bar{2}$
of $SU(2)$ follows from the assignments of $(u\ d)^T$ as a $\bf 2$.

\section{The quantum numbers of the gauge fields}
\noindent
The gauge bosons belong to the adjoint
representation, which in our case is the $SU(6)$'s {\bf 35}.
To identify them, we first decompose the {\bf 35}-dimensional
representation with respect to $SU(3)_C \otimes SU(3) 
\otimes U(1)_{Z'}$ and obtain
\begin{equation}
{\bf 35} = ({\bf 1},{\bf 1})_{0} + ({\bf 8},{\bf 1})_{0} + ({\bf 1},{\bf 8})_{0}
+ ({\bf 3},{\bf \bar{3}})_{2} + ({\bf \bar{3}},{\bf 3})_{-2}
\, . \end{equation}
Secondly, using the branching rules for the {\bf 3}- and 
{\bf 8}-representations of $SU(3)$ into irreps of $SU(2) \otimes U(1)_{Y'}$,
we expand the previous equation as 
\begin{eqnarray}
{\bf 35} &=& ({\bf 1},{\bf 1})_{0,0} + ({\bf 8},{\bf 1})_{0,0} + ({\bf 1},{\bf 1})_{0,0}' + ({\bf 1},{\bf 2})_{3,0} + ({\bf 1},{\bf \bar{2}})_{-3,0} + ({\bf 1},{\bf 3})_{0,0} \nonumber \\
	 & & + ({\bf 3},{\bf 1})_{2,2} + ({\bf 3},{\bf \bar{2}})_{-1,2} + ({\bf \bar{3}},{\bf 1})_{-2,-2} + ({\bf \bar{3}},{\bf 2})_{1,-2}
\, . \end{eqnarray}
Finally we rewrite the $Y'$ and $Z'$ in terms of the new quantum numbers, 
using linear combinations (\ref{TY}) and (\ref{TZ}), for the same 
decomposition as before, arriving at
\begin{eqnarray}
{\bf 35} &=& ({\bf 1},{\bf 1})_{0,0} + ({\bf 8},{\bf 1})_{0,0} + ({\bf 1},{\bf 1})_{0,0}' + ({\bf 1},{\bf 2})_{-1,1} + ({\bf 1},{\bf \bar{2}})_{1,-1} + ({\bf 1},{\bf 3})_{0,0} \nonumber \\
	 & & + ({\bf 3},{\bf 1})_{2/3,1} + ({\bf 3},{\bf \bar{2}})_{5/3,0} + ({\bf \bar{3}},{\bf 1})_{-2/3,-1} + ({\bf \bar{3}},{\bf 2})_{-5/3,0}
\, . \end{eqnarray}
We identify the gauge bosons, as follows: the $({\bf 8},{\bf 1})_{0,0}$ is the 
adjoint representation of $SU(3)_C$, that is, the gluons $G_\mu^i$ 
($i=1,2\ldots 8$); 
the $({\bf 1},{\bf 1})_{0,0}'$ and $({\bf 1},{\bf 3})_{0,0}$ belong to the adjoint
representation of $SU(2)_W \otimes U(1)_W$, and result in the 
bosons of the GWS model, $A_\mu^a$ ($a=1,2,3$) and $B_\mu$;
$({\bf 1},{\bf 2})_{-1,1}$ and its hermitian conjugate (h.c.) are color singlets
and $SU(2)$ doublets, and are the Higgs boson of the GSW model, 
$\widehat{\varphi}$.
The irreps $({\bf 3},{\bf 1})_{2/3,1}$ and 
$({\bf \bar{3}},{\bf 2})_{-5/3,0}$ with their h.c. are the leptoquarks, with 
mixed quantum numbers and thus mediating transitions 
between quarks and leptons. Within our present understanding, there does
not seem to exist, in principle, any particular reason to insist that 
these bosons be either scalar or vector. 
Finally the irrep 
$({\bf 1},{\bf 1})_{0,0}$, with null quantum numbers and representation 
$\propto{\rm diag} (1,1,1,1,1,-5)$ is naturally identified with the Higgs 
whose VEV gives the large-mass GUT scale, since its VEV would not 
produce a vacuum charged in any way. 

With the help of the adjoint constructed as the tensor product of the 
fundamental and its conjugate the gauge bosons can be written out 
in the form
\begin{eqnarray}
\hbox{$\not \! \! A$} + \Phi &=& i \frac{g}{2} \left( \begin{array}{c|cc}%
\hbox{$\bf \not \! \! G$} \cdot \hbox{{\boldmath $\lambda$}} & X & \tilde{X} \\ \hline
X^{\dag} & \hbox{$\bf \not \! \! A$} \cdot \hbox{{\boldmath $\sigma$}} & i \sqrt{2} \gamma^5 \widehat{\varphi} \\
\tilde{X}^{\dag} & i \sqrt{2} {\widehat{\varphi}}^{\dag} \gamma^5 & 0 \\
\end{array} \right) \\%
  & & + i \frac{g}{2} \sqrt{\frac{3}{5}} \hbox{$\not \! \! B$} \left( \begin{array}{c|cc}%
\frac{2}{3} {\bf 1}_{3 \times 3} & 0 & 0 \\ \hline
0 & -{\bf 1}_{2 \times 2} & 0 \\
0 & 0 & 0 \\
\end{array} \right) - \frac{g}{2} \sqrt{\frac{1}{15}} \gamma^5 \; \Upsilon \left( \begin{array}{c|cc}%
{\bf 1}_{3 \times 3} & 0 & 0 \\ \hline
0 & {\bf 1}_{2 \times 2} & 0 \\
0 & 0 & -5 \\
\end{array} \right) \nonumber
\label{DSU6}
\, , \end{eqnarray}
where 
$g$ is the $SU(6)$ coupling constant,
the $\sigma^a$, $a=1,2,3$, are the Pauli matrices,
the $\lambda^i$, $i=1,2\ldots 8$, are the Gell-Mann matrices,
${\bf 1}_{2 \times 2}$ is the $2 \times 2$ unit matrix,
${\bf 1}_{3 \times 3}$ is the $3 \times 3$ unit matrix, and 
\begin{equation}
\widehat{\varphi} = \frac{1}{\sqrt{2}} \left( \begin{array}{c} \varphi^1 - i \varphi^2 \\ \varphi^3 - i \varphi^4 \end{array} \right)
\, , \end{equation}
the GWS Higgs fields.

There is still the open question as to what is the integer spin 
of the leptoquarks. The following argument, apparently inescapable, 
seems to leave no doubt that they must be vector bosons.
To understand the physical behavior of a GYMT one must always 
go back to expansion (\ref{lagexpandido}).
The interactions of the vector bosons among themselves are given by 
the third term on the right of this equation, and the interactions 
of the vector with the scalar bosons are given by the fourth term.
One of the most peculiar facts about GYMT is that 
scalar bosons do not interact among themselves: there is 
no term that does this in the equation. Now, the leptoquarks 
interact with the vector bosons of the Standard Model, so that, 
in order not to contradict its phenomenology, it is necessary 
that they be given a large mass.
But this immediately implies 
that they have to be vector bosons, since, if they were scalars, 
they would not interact with the scalar Higgs. 
In conclusion, in order not to contradict phenomenology, the 
leptoquarks must be vectorial. 

\section{Final comments}
\noindent
It is very satisfactory to see that, after the linear transformation 
of the generators, the quantum numbers of the Standard Model appear 
naturally using the ideas of GYMTs.
This means that not only does the model predict correctly 
all the quantum numbers for both fermions and vector bosons, but that it 
also predicts the correct numbers for {\it both} the GWS and GUT Higgs 
{\it in the same grand unified irrep of the vector bosons}. 

While the GYMT lagrangian is both gauge and Lorentz invariant, a 
gauge transformation of the fermions {\it by itself}  may mix different 
chiralities.
However, transforming at the same time the generalized covariant 
derivative results in an invariant lagrangian.
A maximal subgroup that maintains unchanged the chiral structure of 
the fermion multiplet is a usual Yang-Mills theory in this multiplet.
This is the reason why the GYMTs look like Yang-Mills theories.

There is a detail that does not seem to be working correctly, and has to 
do with the conservation of the number of degrees of freedom before and 
after the GUT symmetry breaking.
The problems is that the GUT Higgs gives mass to 18 leptoquarks, which 
means that there is an increase of 18 dynamical degrees of freedom.
Where do they come from? In the usual GUTs the unitary gauge eliminates 
degrees of freedom from the Higgs 
bosons irrep, but here it is not clear what is happening.

\bibliographystyle{unsrt}

\end{document}